# Valence electron concentration- and N vacancy-induced elasticity in cubic early transition metal nitrides

Soheil Karimi Aghda[a,1], Dimitri Bogdanovski[a], Lukas Löfler[a], Heng Han Sua[a], Lena Patterer[a], Damian M. Holzapfel[a], Arnaud le Febvrier[b], Marcus Hans[a], Daniel Primetzhofer[c], and Jochen M. Schneider[a]

[a] *Materials Chemistry, RWTH Aachen University, Kopernikusstr. 10, 52074 Aachen, Germany*

[b] *Thin Film Physics Division, Department of Physics, Chemistry and Biology (IFM), Linköping University, Linköping SE-58183, Sweden*

[c] *Department of Physics and Astronomy, Uppsala University, Lägerhyddsvägen 1, 75120 Uppsala, Sweden*

## Abstract

Motivated by frequently reported deviations from stoichiometry in cubic transition metal nitride (TMN$_x$) thin films, the effect of N-vacancy concentration on the elastic properties of cubic TiN$_x$, ZrN$_x$, VN$_x$, NbN$_x$, and MoN$_x$ (0.72≤$x$≤1.00) is systematically studied by density functional theory (DFT) calculations. The predictions are validated experimentally for VN$_x$ (0.77≤$x$≤0.97). The DFT results indicate that the elastic behavior of the TMN$_x$ depends on both the N-vacancy concentration and the valence electron concentration (VEC) of the transition metal: While TiN$_x$ and ZrN$_x$ exhibit vacancy-induced reductions in elastic modulus, VN$_x$ and NbN$_x$ show an increase. These trends can be rationalized by considering vacancy-induced changes in elastic anisotropy and bonding. While introduction of N-vacancies in TiN$_x$ results in a significant reduction of elastic modulus along all directions and a lower average bond strength of Ti–N, the vacancy-induced reduction in [001] direction of VN$_x$ is overcompensated by the higher stiffness along [011] and [111] directions, resulting in a higher average bond strength of V–N. To validate the predicted vacancy-induced changes in elasticity experimentally, close-to-single-crystal VN$_x$ (0.77≤$x$≤0.97) are grown on MgO(001) substrates. As the N-content is reduced, the relaxed lattice parameter $a_0$, as probed by X-ray diffraction, decreases from 4.128 Å to 4.096 Å. This reduction in lattice parameter is accompanied by an anomalous 11% increase in elastic modulus, as determined by nanoindentation. As the experimental data agree with the predictions, the elasticity enhancement in VN$_x$ upon N-vacancy formation can be understood based on the concomitant changes in elastic anisotropy and bonding.

---
[1] Corresponding author

**Introduction**

When it comes to material systems suitable for hard protective coatings [1, 2] and diffusion barrier layers [3, 4], group IV, V, and VI transition metal nitrides, or TMNs, are often the materials of choice in industrial applications, due to their excellent thermal stability [1, 5], corrosion resistance [6], and mechanical properties [7, 8]. Additionally, the properties of binary TMN compounds can be enhanced by modifying the occupancy of metal and non-metal sublattices via alloying constituents. For example, metastable ternary (Ti,Al)N [9] shows age hardening at elevated temperatures before decomposing into its stable constituents [10], and metastable quaternary (Ti,Al)(O,N) exhibits significant thermal stability enhancement in comparison to the (Ti,Al)N counterpart [11].

TMNs usually crystallize in a cubic structure (space group $Fm\bar{3}m$, NaCl prototype) for group IV [12-14], V [15-17], and VI [18-20] elements. Most of these binary compounds exhibit a large single-phase field over a significant composition range, hence, the accommodation of point defects can be expected. Formation and stability of point defects such as vacancies [21-24], interstitials [22, 25], and Frenkel defects [21, 23, 26] in TMNs and (TM,Al)Ns have recently been investigated. In general, point defects have been shown to exert great influence on the stability and mechanical properties of the nitride compounds. For instance, vacancies as the lowest-energy defects [22] stabilize the mechanically unstable cubic TaN [27], MoN [27, 28], and WN [29], and enhance hardness in TiN [30, 31], toughness in $V_{0.5}Mo_{0.5}N$ [32], and elastic modulus in NbN [16].

With respect to elasticity, vacancy-induced changes in the elastic modulus seem to follow a different trend for early TMN compounds. For $TiN_x$ ($x$ = N/Ti) [31, 33], an increase in N vacancy concentration, from $TiN_{1.00}$ to $TiN_{0.67}$, has been experimentally shown to continuously reduce the elastic modulus, from ~428 to 325 GPa, respectively. The vacancy-induced reduction in elasticity of $TiN_x$ was attributed to a reduction in the bond density of the compound [31]. Similar behavior is also reported for $HfN_x$ [34], another compound from group IV TMNs. However, reports on group V NbN$_x$ indicated an anomalous increase in elastic modulus via both N and Nb vacancy incorporation, as observed both theoretically [16] and experimentally [35]. For TaN$_x$ thin films [36], a reduction in elastic modulus is measured as $x$ is increased from stoichiometric $x = 1.00$ to overstoichiometric $x = 1.35$, which is the opposite of the behavior observed for NbN$_x$, another group V TMN [16]. Apart from the scattered literature on the vacancy-induced changes in mechanical properties of TMNs, there is no systematic study on the effect of vacancies in conjunction with the increase in valence electron concentration (VEC) on the elastic properties of group IV to V and VI TMNs.

Vivid morphological differences, which are typically not considered in the structural models used in *ab initio* simulations, make correlative experimental and theoretical studies challenging with respect to elastic properties. In order to deconvolute the intrinsic properties of the material systems from morphological effects such as grain bouondaries, an effective approach is to carry out property measurements on epitaxial close-to-single-crystal layers, as previously done for ScN(001) [37], TiN(001) [4, 38], HfN(001) [14], VN(001) [39], NbN(001) [16], and CrN(001) [18], to name a few.

Here, we systematically investigate the effect of N vacancy concentration on the elastic properties of binary cubic TiN$_x$, ZrN$_x$, VN$_x$, NbN$_x$, and MoN$_x$ ($0.72 \leq x \leq 1.00$) by density functional theory (DFT) calculations. The predictions are validated experimentally for epitaxially grown VN$_x$(001) ($0.77 \leq x \leq 0.97$) thin films. The calculated bulk moduli, Poisson's ratios, and elastic moduli of the binary compounds with varying N vacancy concentration exhibit dependency on the VEC in groups IV, V, and VI. It is demonstrated that while for TiN$_x$ an increase in N vacancy concentration results in a reduction in elastic modulus, a N vacancy-induced elasticity enhancement is observed in the VN$_x$. The results of crystal orbital Hamilton population (COHP) analyses and elastic anisotropy calculations reveal that the stiffness increase (decrease) upon vacancy introduction into $VN_x$ ($TiN_x$) is caused by direction-dependent bond-strengthening (bond-softening).

**Computational details**

*Ab initio* calculations were performed using density functional theory (DFT) [40, 41] as implemented in the Vienna *ab initio* Simulation Package (VASP, version 5.4.4, University of Vienna) [42-44], employing projector-augmented waves (PAW) [45, 46] with a cut-off energy of 500 eV for basis set representation. The electronic exchange-correlation energy was calculated with the well-established parametrization of the generalized gradient approximation (GGA) by Perdew, Burke, and Ernzerhof (PBE) [47]. A Γ-centered k-point mesh with dimensions of 9 × 9 × 9 constructed via the Monkhorst-Pack method [48] was used to sample the Brillouin zone, using the Methfessel-Paxton method for integration [49]. For all *ab initio* simulations, 2 × 2 × 2 supercells with cubic B1 symmetry containing 64 atoms were constructed for TiN, ZrN, VN, NbN, and MoN. Three (corresponding to 4.7%), six (9.4%), or nine (14.1%) vacancies on random positions of the N sublattice were then introduced, yielding a total of 20 different structures. The stiffness tensors of these systems were determined with the strain-stress method as implemented in Ref. [50] and projected onto the cubic symmetry as described in the literature [51]. From the stiffness tensor, the bulk, shear, and elastic modulus, as well as Poisson's ratio, were calculated with Hill's approximation [52]. Additionally, temperature-dependent equilibrium volume and elastic moduli were calculated by the Debye-Grüneisen model [53] as described elsewhere [54]. Furthermore, the directional elastic moduli were calculated with the help of equation (1), as derived by Nye [55].

$$\frac{1}{E_{hkl}^{cub}} = S_{11} - 2(S_{11} - S_{12} - \frac{1}{2}S_{44})(\bar{h}^2\bar{k}^2 + \bar{k}^2\bar{l}^2 + \bar{h}^2\bar{l}^2) \qquad (1)$$

with $[\bar{h}\bar{k}\bar{l}]$ as the normalized vector along the direction $[hkl]$ and $S_{ij}$ as elements from the compliance tensor (S).

In addition to the elastic properties, the electronic structure and bonding characteristics were analyzed for two of the systems, $VN_x$ and $TiN_x$, with an ideal vacancy-free structure and the three N vacancy concentrations described above. The optimized structures used for the calculation of elastic properties were utilized as input structures. Single-point (static) DFT simulations using VASP were performed, employing a slightly reduced k-mesh of 5 × 5 × 5 and the Blöchl tetrahedron method for Brillouin zone integration [56], with other key settings similar to the preceding simulation series described above. The valence electron configurations of the PAW potential files were $3p^6 4s^1 3d^4$ for V, $3s^2 3p^6 4s^1 3d^3$ for Ti and $2s^2 2p^3$ for N. The wavefunctions of the simulated systems were generated by VASP and post-processed with the LOBSTER package (version 4.0.0, Institute of Inorganic Chemistry, RWTH Aachen University) [57-60] in order to project the delocalized, plane-wave-based information onto local orbitals. This enables the calculation of a more precise, atom- and orbital-resolved density of states, as well as the (integrated) crystal orbital Hamilton population ([I]COHP) [61] in order to estimate the bonding character and strength of individual interatomic bonds. It should be noted that, while the COHP is not a *direct* descriptor of the bond strength, it is strongly correlated with the latter property and is routinely used to assess it in various systems [62-65].

**Experimental details**

Vanadium nitride ($VN_x$) thin films from stoichiometric to understoichiometric were synthesized using reactive direct current magnetron sputtering (DCMS) in a load-locked ultra-high-vacuum chamber. The base pressure of the system was below $1 \times 10^{-5}$ Pa at respective deposition temperatures ($T_s$) of 230, 430, 600, and 700 °C. An elemental V target (> 99.5% purity) with a diameter of 50 mm was powered in DC at a constant averaged-power of 100 W, which resulted in a power density of ~5 W/cm$^2$. Initial growth experiments performed in Ar/$N_2$ working gas mixture (not shown here) resulted in extremely N-deficient thin films, hence, near-stoichiometric cubic $VN_x$ could not be achieved even at the lowest $T_s$ = 230 °C employed here. This has also been reported by Mei *et al.* [39], where they systematically investigated the effect of $N_2$ partial pressure on the chemical composition of $VN_x$. In addition, high energetic $Ar^+$ irradiation-induced defects [23, 66, 67] could hinder the growth of high crystalline quality, epitaxial $VN_x$ thin films. Therefore, the working gas was pure $N_2$ (5.0 purity) with a deposition pressure of 2.6 Pa. Such a high discharge pressure was employed to thermalize sputtered atoms and neutralize the majority of ions [68].

MgO(001) substrates with a dimension of $10 \times 10 \times 0.5$ mm$^3$ were mounted at a defined target-to-substrate distance of 6 cm. Prior to the depositions, the unpolished backsides of MgO(001) substrates were coated with TiN to optimize heat conduction and avoid localized heating effects. Moreover, in order to optimize the surface of MgO(001) substrates, wet-cleaning was conducted by using isopropanol and acetone in sonication [69], accompanied by a $N_2$ blow-drying procedure. Lastly, an annealing step was performed at 800 °C for one hour in the deposition chamber prior to the deposition of the desired thin films. The substrate holder was kept at floating potential for all depositions and the thin films were grown to a thickness of approximately ~400 nm.

Time-of-flight elastic recoil detection analysis (ToF-ERDA) at the Tandem Laboratory of Uppsala University [70] was used for depth profiling of the film composition. A primary ion beam of 36 MeV $^{127}$I$^{8+}$ was employed and the detection telescope, including a solid-state detector was located at 45° with respect to the primary ion beam. The incident and exit angle of the ions and detected recoils with respect to the specimen surface were both 22.5°. Depth profiles were obtained from time-energy coincidence spectra by using CONTES [71] and a TiN reference sample [72], which has been characterized by Rutherford backscattering spectrometry, was also probed. All resulting depth profiles were found homogeneous and the maximum oxygen impurity content was 1 at.%. The total maximum measurement uncertainties were 3% relative deviation of the deduced values for V and N.

For the X-ray photoelectron spectroscopy (XPS) measurements, samples were inserted in an AXIS Supra instrument (Kratos Analytical Ltd.) equipped with a monochromatic Al-K$_\alpha$ X-ray source. The base pressure of the system during acquisition was < 5.0 × 10$^{-6}$ Pa. High-resolution N 1s spectra were obtained using a pass energy of 10 eV and a step size of 0.04 eV (6 sweeps, dwell time 1000 ms). The measurement spot size was 700 × 300 µm$^2$. The binding energy (BE) scale of the spectrometer was calibrated using a sputter-cleaned Ag standard (Ag 3$d_{5/2}$ signal at 368.2 eV). No charging effect of the VN$_x$ samples was observed.

Structural analysis of the thin films was performed with a Siemens D5000 X-ray diffraction (XRD) system (Munich, Germany) using a Cu K$_\alpha$ radiation source, operated at a voltage and current of 40 kV and 40 mA, respectively. The X-ray source and the detector were coupled in *θ-2θ* scans (Bragg-Brentano geometry), scanning a *2θ* range from 41° to 46° to obtain the (200) diffractions from VN$_x$ thin films and the MgO substrate. A step size of 0.05° and dwell time of 2 s per step were used for the measurements. *θ*-rocking curves along (200) diffraction plane of VN$_x$ were acquired with an incident parallel beam within the same diffractometer.

Reciprocal space maps (RSM) were acquired using a PANalytical Empyrean diffractometer with Cu Kα radiation for the phase structure and growth orientation analysis. Symmetric and asymmetric RSMs were recorded around MgO(002) and MgO(204), respectively using a four-axis goniometer and a primary optics consisting of a parabolic graded multilayer mirror, collimator, and a channel-cut 2-bounce Ge(220) monochromator. The in-plane coherence lengths ($\xi_\parallel$), corresponding to average mosaic domain sizes [73], are determined from the widths of (002) diffraction peaks parallel to the diffraction vector [74]:

$$\xi_\parallel = \frac{2\pi}{\Delta q_x} = \frac{\lambda}{2\varGamma_\theta \sin\theta} \qquad (2)$$

where $\varGamma_\theta$ is the full-width at half-maximum of the peak intensities of the VNx (002) diffractions along the $\theta$ direction.

The surface topography and morphology of the thin films were characterized using scanning electron microscopy (SEM) at an acceleration voltage of 10 kV and a current of 50 pA within an FEI Helios Nanolab 660 dual-beam microscope (Hillsboro, OR, USA).

The elastic modulus $E$ was determined by nanoindentation using a Hysitron (Minneapolis, MN, USA) TI-900 TriboIndenter equipped with a Berkovich geometry diamond tip with 100 nm radius. At least 25 quasistatic indents with a maximum load of 0.8 mN were performed. Indentation depths were less than 40 nm, which is ~5% of the film thickness, to minimize the substrate effect [75]. The tip area function was determined with a fused silica reference before each measurement series and verified to remain unchanged thereafter. The reduced modulus was acquired from the unloading segment of load-displacement curves using the method of Oliver and Pharr [76]. The elastic moduli of the films were then obtained from the reduced moduli data using the composition-dependent Poisson's ratios calculated in this study and the isotropic approximation.

**Results and discussion**

The calculated bulk moduli, together with the Poisson's ratios, and elastic moduli of the binary nitrides in the ground state, as a function of the N/TM ratio *x* are depicted in Figure 1 (a-c). The introduction of N vacancies in the cubic binary $TMN_x$ results in a reduction of the bulk modulus, independent of the TM element, see Figure 1 (a). The decrease is ~17% for group IV nitrides ($TiN_x$ and $ZrN_x$), as *x* changes from 1.00 to 0.72. A less significant vacancy-induced reduction in bulk modulus of ~9% is calculated for the group V nitrides ($VN_x$ and $NbN_x$). Moreover, for the group VI binary $MoN_x$, the stoichiometric cubic structure is not mechanically stable, as already reported by Balasubramanian *et al.* [28]. However, it has been shown that the introduction of N vacancies results in the stabilization of the cubic understoichiometric $MoN_x$. The influence of N vacancy concentration on the bulk modulus for $MoN_x$ is significantly lower than for the other nitrides, with a reduction of ~2% from $MoN_{0.91}$ to $MoN_{0.72}$. These results indicate that increasing VEC in the cubic binary $TMN_x$ leads to a smaller N vacancy-induced variation in bulk modulus for the cubic binary compounds.

The calculated N vacancy-dependent Poisson's ratios for the binary $TMN_x$ are shown in Figure 1 (b). Here, the Poisson's ratio values follow completely different trends with respect to the N vacancy concentration for each of the group IV, V, and VI nitrides. While for $TiN_x$ and $ZrN_x$, the Poisson's ratio increases with increasing N vacancy concentration, there is an opposite trend for $VN_x$ and $NbN_x$ and a nearly N vacancy-concentration-independent Poisson's ratio is calculated for $MoN_x$. As the elastic modulus $E$ is linked to the bulk modulus $B$ and Poisson's ratio $v$ via the relationship $E = 3 \times B \times (1 - 2v)$ within the isotropic approximation, the concomitant changes in both bulk modulus and Poisson's ratio are consistent with the elastic modulus data in Figure 1 (c). For group IV nitrides, we observe a continuous reduction in $E$ with respect to N vacancy concentration. Contrary, $VN_x$ shows an anomalous increase in elastic modulus due to the presence of N vacancies, up to 10% from $VN_{1.00}$ to $VN_{0.72}$. For $NbN_x$, which has the same VEC as $VN_x$, there is also a slight, albeit much less pronounced, increase in elastic modulus as the N vacancy are introduced. This behavior is also present in the case of $MoN_x$.

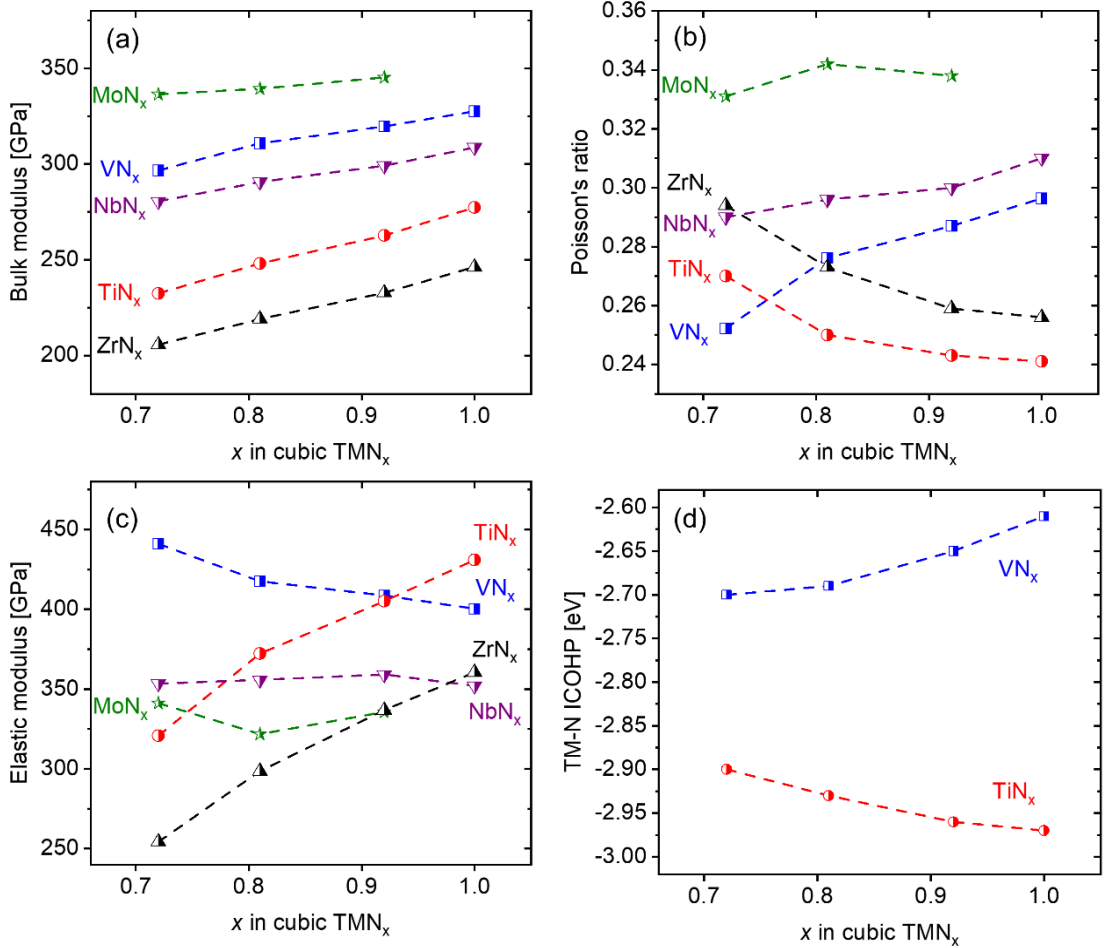

*Figure 1: (a) Bulk modulus, (b) Poisson's ratio, and (c) elastic modulus at ground state for cubic $TMN_x$ (TM = Ti, V, Zr, Nb, and Mo) as well as (d) the average ICOHP v alues determined for TM-N bonds in cubic $TiN_x$ and $VN_x$ as a function of x.*

As the elastic modulus of binary TMN is known to be anisotropic [77], different directional elastic responses for $TiN_x$ and $VN_x$, with and without vacancies, were analyzed. The results are depicted in Figure 2 showing the elastic modulus in the three main planes, (001), (011), and (111). In both stoichiometric compounds the directional elastic modulus is the strongest along the [100] direction, see Figure 2 (a) and (d), consistent with [77]. The introduction of N vacancies results in a significant reduction of the direction-dependent $TiN_x$ elastic modulus, see Figure 2 (a), (b), and (c). In contrast, for $VN_x$, the vacancy-induced reduction in elastic modulus along the [100] direction is smaller and overcompensated by concomitant changes along the [110] and [111] directions. Therefore, as vacancies are introduced in TiN$_x$, marginal changes in directionality are accompanied by a reduced elastic modulus in all directions, while in VN$_x$, N vacancies cause changes in the directional elastic moduli leading to an overall increase in stiffness.

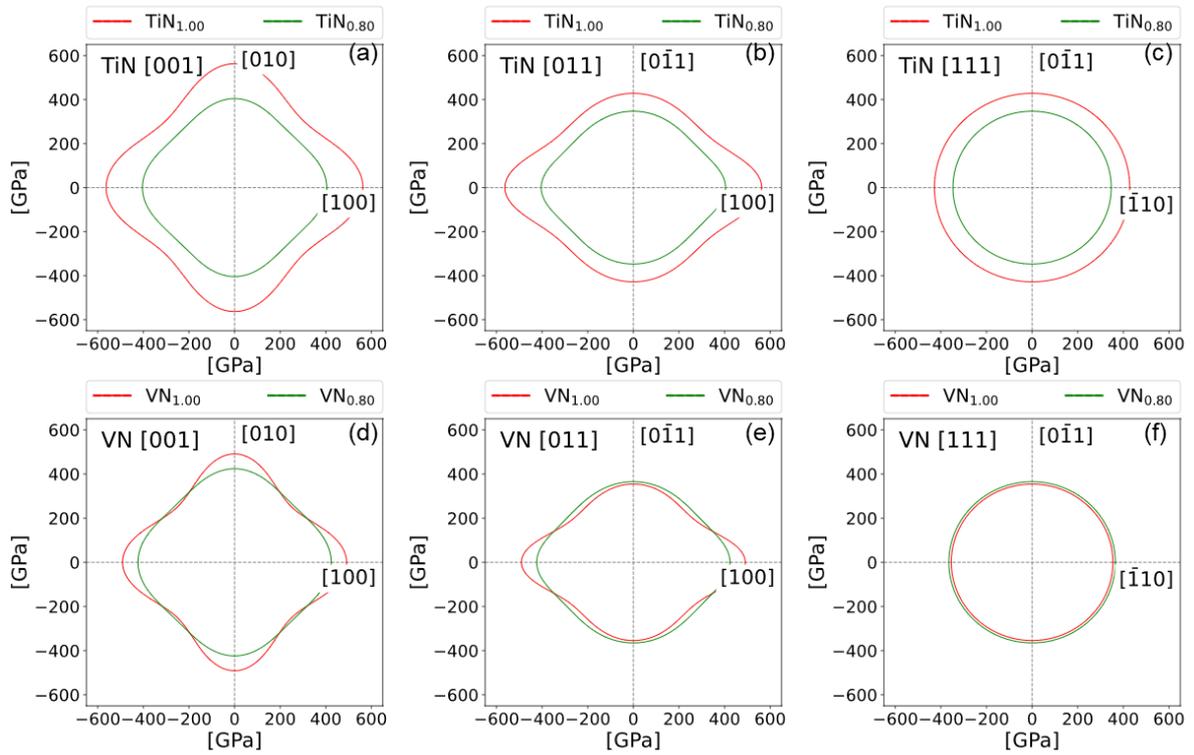

*Figure 2. Directional elastic modulus for the three main projection planes, [001], [011], and [111] in (top) TiN$_x$ and (bottom) VN$_x$.*

To identify the cause for the different N vacancy-induced elastic modulus changes between group IV and V nitrides, total and partial densities of states (DOS) are calculated for TiN$_x$ and VN$_x$, with the results shown in **Error! Reference source not found.** of the supplementary material. It is evident that there is a qualitative similarity regarding the general structure of the DOS between stoichiometric TiN and VN. However, in the proximity of the Fermi level ($E_f$), the VN DOS exhibits local maxima, which indicates localized states accounting for the electronic instability. Additionally, introduction of a N vacancy in VN$_x$ results in slightly fewer occupied states near $E_F$, which indicates that N vacancies contribute to an electronic stabilization of the cubic structure of VN$_x$ [78]. For both TiN$_x$ and VN$_x$, the emergence of peaks at –2 eV is attributed to the vacancy-derived states, in good agreement with other reported electronic structures for the understoichiometric cubic TiN$_x$ [33] and VN$_x$ [78]. Nonetheless, there is no other significant change in the DOS between the stoichiometric and understoichiometric compositions, which is why we will in the following focus on the differences in local electronic structure in the vicinity of a N vacancy and compare the DOS to that of the pristine non-vacancy-containing system. The results of total and partial differential density of states (dDOS) are illustrated in Figure 3 (a) and (b) for TiN$_x$ and VN$_x$, respectively. The electronic structure of both TMN$_x$ compounds is dominated by TM 3$d$-N 2$p$ overlap, which indicates $sp^3d^2$ hybridization. While the introduction of N vacancies in TiN$_x$ resulted in reduced hybridization between Ti and N within the energy range −4 eV to −9 eV (Figure 3 (a)), no significant changes between VN$_{1.00}$ and VN$_{0.81}$ can be observed in Figure 3 (b) in this energy range. These results indicate that, unlike TiN$_x$, the strong covalent bonding nature of VN$_x$ remains unchanged as N vacancies are introduced. These local effects, however, are diluted in the total DOS and could not be resolved by considering the entire supercells.

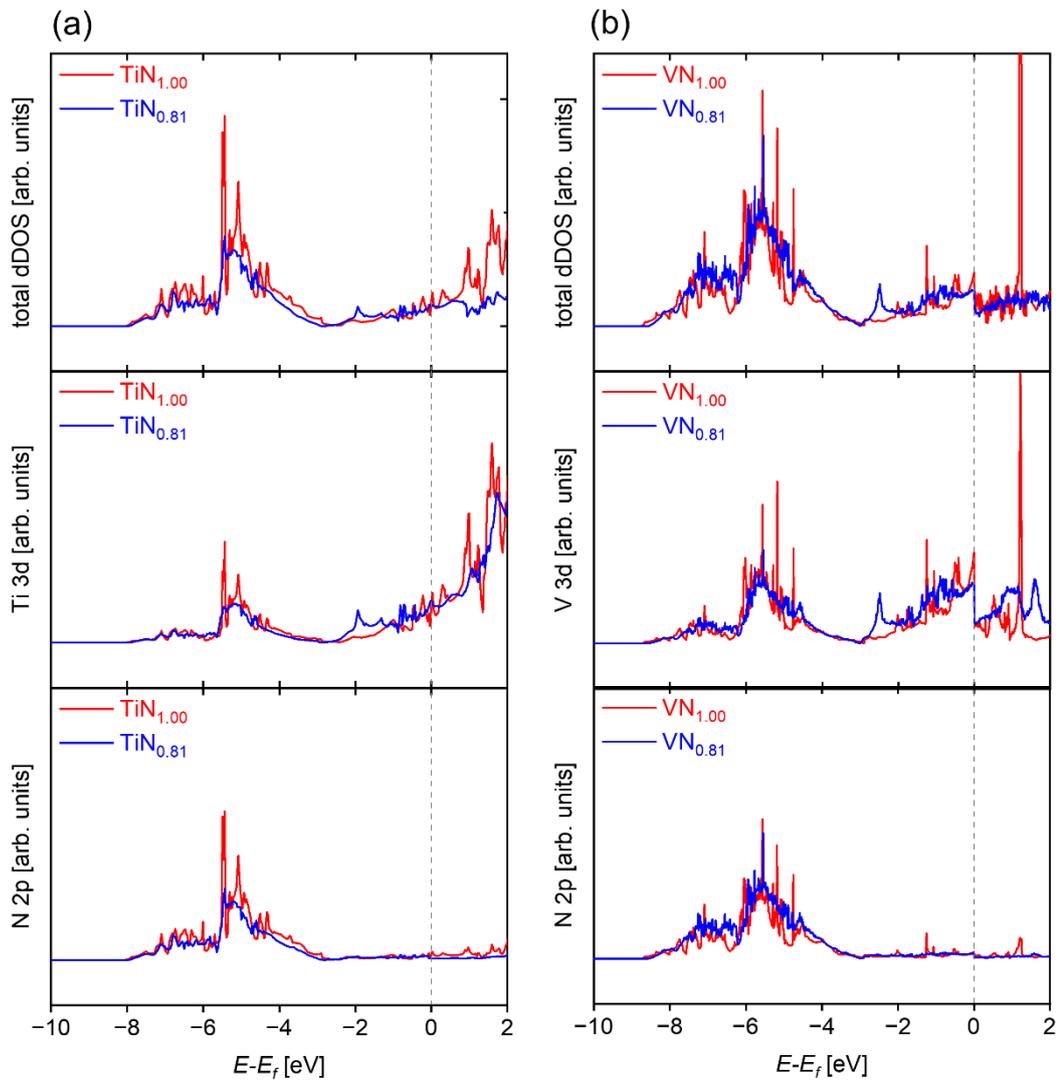

*Figure 3. Total and partial differential density of states (dDOS) analysis in the vicinity of a N vacancy for the vacancy-containing systems and a similar region in the fully-occupied systems for (a) TiN$_x$ and (b) VN$_x$ as a function of x. $E_f$ designates the Fermi energy.*

To quantitatively evaluate the bond strength of the binary compounds with and without the presence of N vacancies, we will focus on the crystal orbital Hamilton population (COHP) analyses. The average bond energies for TM–N bonds in TiN$_x$ and VN$_x$ with respect to the N vacancy concentration are obtained from COHP calculations, where the integrated COHP (ICOHP) for these two binary nitrides is plotted in Figure 1 (d). Interestingly, the trends of the ICOHP are completely different for TiN$_x$ and VN$_x$ with respect to N vacancy concentrations. While increasing N vacancy concentration in TiN$_x$ results in a more positive ICOHP value, correlated with a weakening of the bond, a more negative ICOHP for VN$_x$ with respect to increasing N vacancy concentration suggests that bond strengthening seems to be the primary reason behind the anomalous stiffness increase. These findings are consistent with the calculated data by Rueß *et al.* [54], who showed that V vacancy-induced bond strengthening is the origin of the stiffness increase in overstoichiometric VN$_x$.

Another important aspect of the presence of N vacancies in cubic binary TMN$_x$ is the local lattice relaxation. Therefore, we systematically analyze changes in the bonding character with respect to N vacancy concentration for both TiN$_x$ and VN$_x$, with the results summarized in Figure 4. Figure 4 (a) and (d) show the local lattice relaxations evident in the structural model, projected in the (100) plane, which are induced by the presence of a N vacancy in TiN$_{0.81}$ and VN$_{0.81}$, respectively. For TiN$_{0.81}$ the distance between two Ti atoms along [100] is increased by ca. 2.6% from 4.24 Å in pristine TiN to 4.35 Å near the vacancy site, which shows an outward displacement of the atoms. However, the local relaxation along the vacancy site of $VN_{0.81}$ shows an inward displacement of V atoms from 4.13 Å to 3.86 Å by 6.5% towards the vacancy site. Figure 4 (b), (c), (e), and (f) depict the effect of the local lattice relaxations upon bond length and bond strength (expressed via the ICOHP) distributions within $TiN_x$ and $VN_x$ as a function of the vacancy concentration. For both binary compounds, the introduction of a N vacancy leads to a distribution in the overall bond lengths and energies, in contrast to the stoichiometric systems. However, the vacancy-induced distribution is much larger in the case of $VN_x$, as evident from the scattering from Figure 4 (e). The bond strength histograms in Figure 4 (c) and (f) illustrate the changes in the TM–N bond energy with respect to N vacancy concentration in $TiN_x$ and $VN_x$, respectively. It is evident that for $TiN_x$ the average bond energy (ICOHP) becomes more positive as the N vacancy concentration is increased, going from –2.97 for TiN to –2.90 eV for TiN$_{0.72}$. Thus, a majority of the bonds are weaker in comparison to the average bond in stoichiometric TiN, representing vacancy-induced softening. In contrast, most of the bonds in understoichiometric VN$_x$ have lower energies than in the stoichiometric compound, as exhibited in Figure 4 (f). Here, a decrease in the ICOHP from –2.62 for VN to –2.70 eV for VN$_{0.72}$ indicates that the vacancy-induced changes in elastic anisotropy, Figure 2, are directly connected to overall vacancy-induced bond strengthening in VN$_x$.

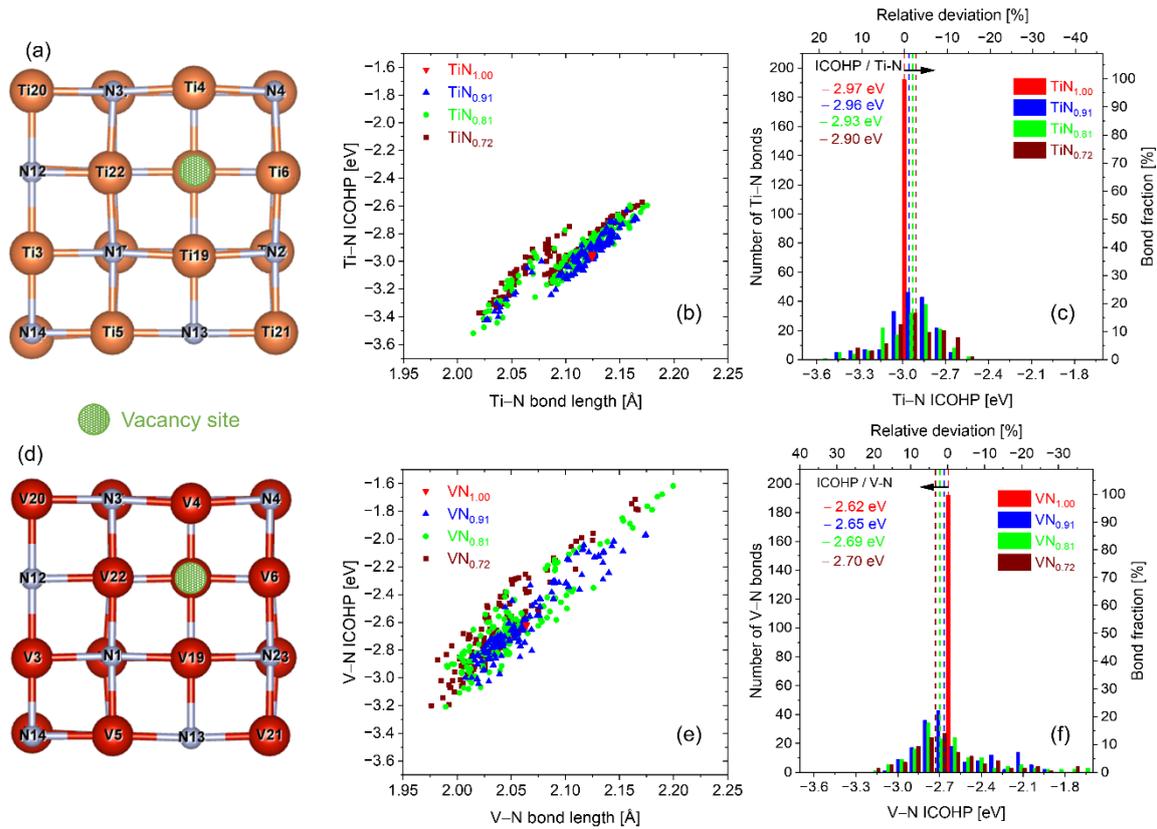

*Figure 4. Projections of the structural models in the (001) plane, depicting the lattice close to the vacancy site for (a) $TiN_{0.81}$ and (d) $VN_{0.81}$. (b) Ti-N ICOHP values vs. Ti-N bond length. (c) Ti-N ICOHP histogram. (e) V-N ICOHP values vs. V-N bond length. (f) V-N ICOHP histogram. The average ICOHP for the TM-N bond of each structure is indicated with dashed lines and the corresponding average values are included in figures (c) and (f).*

In an effort to experimentally validate the N vacancy-induced bond strengthening, VN$_x$ thin films were grown epitaxially on single crystal MgO(001) substrates as a function of $T_s$ in a pure N$_2$ atmosphere. The resulting thin film compositions, as obtained by ToF-ERDA, are plotted in Figure 5 (a), showing the N/V ratio $x$ with respect to $T_s$. A steep reduction in $x$ is evident as a function of substrate temperature, where an increase from 230 to 700 °C leads to a reduction in the $x$ value from near-stoichiometric 0.97 ± 0.06 to understoichiometric 0.77 ± 0.04. The reduction in the N content of the layers at higher temperatures is evidently attributed to N$_2$ desorption due to the thermally-activated mechanisms on the surface of the growing film, as has been shown theoretically by Sangiovanni *et al.* [79]. It can be also learned from Figure 5 (a), that near-stoichiometric VN(001) thin films can be synthesized at temperatures below ~430 °C under the here stated deposition conditions, as also reported by Mei *et al.* [80].

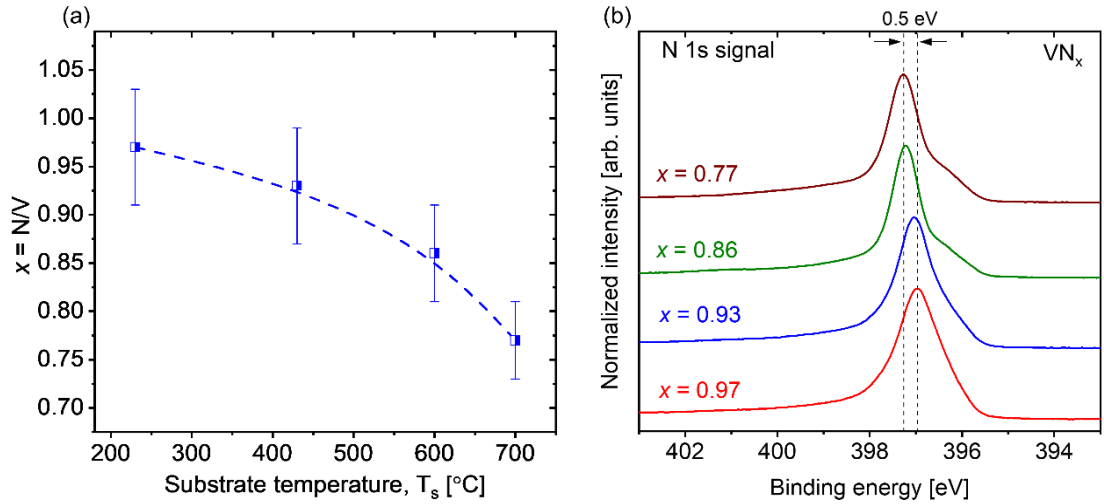

*Figure 5. (a) N/V ratio x of VN$_x$/MgO(001) thin films deposited by DCMS as a function of substrate temperature determined by TOF-ERDA. (b) XPS N 1s spectra of (a) VN$_x$/MgO(001) with respect to N/V ratio x.*

The presence of N vacancies has been further indicated by XPS measurements, with the N 1$s$ spectrum shown for the VN$_x$ ($0.86 \leq x \leq 0.97$) thin films in Figure 5 (b). The measured N 1$s$ spectrum for the VN$_{0.97}$ thin film exhibits a main peak centered at ~397.0 eV. Upon decreasing $x$ from 0.97 to 0.86, the main N 1$s$ binding energy (BE) continuously shifts towards higher values by approximately 0.5 eV, up to ~397.5 eV. The appearance of an additional N 1$s$ feature (BE ~396 eV) in samples with lower $x$ ($\leq 0.86$) is mainly correlated with the oxidized state [81]. The shift in BE with respect to $x$ has also been observed for other transition metal nitrides and can be understood in terms of vacancy-induced changes in charge distributions [81-83].

XRD $\theta$-$2\theta$ diffractograms over the $2\theta$ range from 41 to 46°, across 002 Bragg diffraction peaks, obtained from VN$_x$/MgO(001) thin films with $0.86 \leq x \leq 0.97$ together with a plain MgO(001) substrate are shown in Figure 6 (a). The MgO(002) diffraction peak is positioned at $2\theta$ = ~43°. The (002) diffraction peak intensity of the deposited VN$_x$ thin films increases with higher $T_s$, accompanied by a N vacancy-induced monotonic shift from 43.4° at $T_s$ = 230 °C and $x$ = 0.97 to 44.2° at $T_s$ =700 °C and $x$ = 0.77. This peak shift towards larger diffraction angles accounts for a smaller lattice parameter due to an increase in N vacancy concentration, which is consistent with the compositional data exhibited in Figure 5 (a). In order to evaluate the crystalline quality of the grown thin films, 002) $\theta$-rocking curves for $VN_x$ thin films were measured and are depicted in Figure 6 (b). The $VN_x$ thin films deposited at $T_s$ = 230, 430, 600, and 700 °C exhibit a full-width-at-half-maximum (FWHM) value of $\Gamma_\theta$ = 2.02, 1.67, 0.22, and 0.24°, respectively. These results indicate that the increase in $T_s$ up to 600 °C leads to a reduction in the mosaicity or enhancement in the preferentially-oriented $VN_x$ thin films. However, further increase in $T_s$ up to 700 °C, and subsequent rise in N vacancy concentration promote missorientation of the $VN_x$ crystallites. The significant influence of the deposition temperature upon the crystal structure evolution of the thin films has direct consequences for the morphological evolution, as probed by surface SEM imaging in Figure 6 (c-f). The presence of the faceted structures within the (002) matrix of the $VN_{0.97}$ deposited at low $T_s$ (230 °C) can be correlated with limited adatom surface mobility [84]. On the other hand, very smooth featureless surfaces have been obtained as $T_s$ is increased to above 600 °C, see Figure 6 (e) and (f).

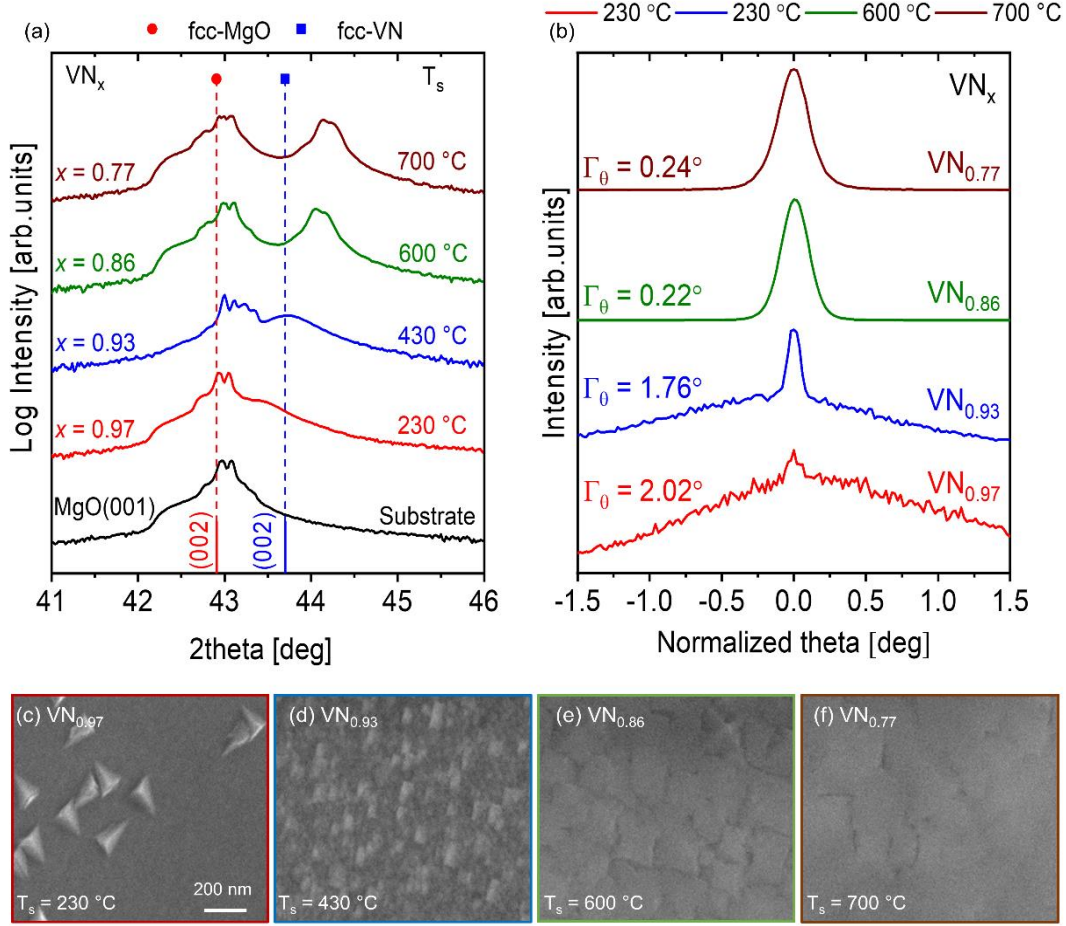

*Figure 6. (a) XRD θ-2θ diffractograms, (b) (002) θ-rocking curves with calculated FWHM ($\Gamma_\theta$), and (c) surface SEM micrographs of VN$_x$/MgO(001) thin films as a function of substrate temperature. The N/V ratio x for each thin film is shown.*

The results of high-resolution XRD reciprocal space maps (HR-RSM) of the thin films grown at $T_s$ = 430, 600, and 700 °C with $x$ = 0.93, 0.86, and 0.77, respectively, acquired over symmetric 002 and asymmetric 204 reflections are shown in Figure 7. Moreover, the in-plane coherence length ($\xi_\parallel$), in-plane strain ($\varepsilon_\parallel$), and relaxed lattice parameter ($a_0$) were determined as a function of $T_s$ from the HR-RSM results and are listed in Table 1.

At $T_s$ = 430 °C, a significant broadening is observed for the in-plane direction ($q_x$) along the *θ-2θ* direction is observed, see Figure 7 (a), which accounts for a large mosaicity in this sample. The mosaicity is significantly reduced at higher deposition temperatures, as denoted by a smaller in-plane diffraction broadening, Figure 7 (b) and (c). In-plane coherence length $\xi_\parallel$ increases from 17 to 93 nm as $T_s$ is increased from 230 to 600 °C, respectively. The here reported $\xi_\parallel$ = 93 nm for understoichiometric VN$_{0.86}$ is among the largest in-plane coherency lengths obtained for the epitaxially grown binary transition-metal nitrides reported so far [13, 18, 34, 36-39], which reflects the high crystalline quality of this thin film. Further increase in $T_S$ to 700 °C for the growth of $VN_{0.77}$ resulted in a reduction of $\xi_\parallel$ to 62 nm, which can be correlated to a higher concentration of N vacancies in this highly understoichiometric film [39].

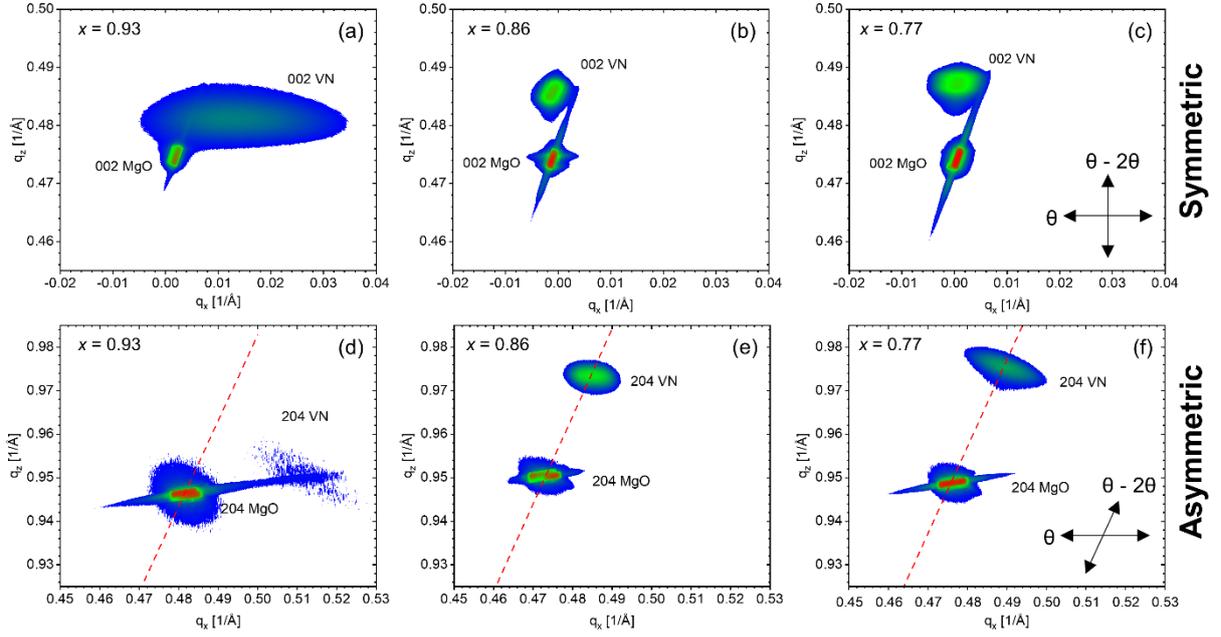

*Figure 7. Symmetric (top) and asymmetric (bottom) HR-RSMs of DCMS-deposited $VN_x$ thin films at (a) & (d) 430 °C, (b) & (e) 600 °C, and (c) & (f) 700 °C. Dashed red lines in the asymmetric maps extend from the origin, located outside the panels, along the (204) direction.*

*Table 1: $x$ ratio, in-plane coherence lengths $\xi_\parallel$, in-plane strain $\varepsilon_\parallel$, relaxed lattice parameter $a_0$ of $VN_x$/MgO(001) thin films at various substrate temperatures.*

| VN$_x$/MgO(001) | | | | |
|---|---|---|---|---|
| $T_S$ [°C] | $x$ | $\xi_\parallel$ (nm) | $\varepsilon_\parallel$ (%) | $a_0$ (Å) |
| 230 | 0.97 | 17 | −2.01 | 4.128 |
| 430 | 0.93 | 22 | −1.46 | 4.119 |
| 600 | 0.86 | 93 | −0.51 | 4.099 |
| 700 | 0.77 | 62 | −0.18 | 4.096 |

Furthermore, from the asymmetric maps, a fully-relaxed thin film on the single crystal substrate has the 204 diffraction peaks centered on the dashed line extending from the origin along the 204 diffraction peaks of the substrate. Here, the lateral displacement of the 204 diffraction peaks of the film deposited at $T_s$ = 430 °C to the right side of the line indicates the presence of compressive strain with an in-plane biaxial strain $\varepsilon_\parallel$ of −1.46% in this layer. A slightly higher $\varepsilon_\parallel$ = −2.01% has been measured for the $VN_{0.97}$ film deposited at 230 °C. However, almost fully-relaxed thin films are obtained at $T_s$ = 600, and 700 °C with $\varepsilon_\parallel$ of −0.51 and −0.18, respectively.

Relaxed lattice parameters $a_0(x)$ obtained from the HR-RSMs are summarized in Table 1. The $a_0(x)$ decreases from 4.128 Å for $VN_{0.97}$ to 4.096 Å for $VN_{0.77}$, following a nearly linear trend. This negative slope with respect to $x$ is also observed in other works for $VN_x$ [39] and $TiN_x$ [31]. The relaxed lattice parameters from the thin films are plotted in Figure 8 (a) together with the lattice parameters obtained from the DFT calculations by considering N vacancies as the origin of understoichiometry in $VN_x$. The maximum deviations between the calculated and measured values are below 1.0%, signifying good agreement. In general, the calculated lattice parameters are found to be smaller than the experimentally measured values, however, showing a similar negative trend with respect to $x$. The good agreement between the experimental data and the predictions in turn suggests that the formation of understoichiometric $VN_x$ is governed by N vacancies.

By considering the high crystalline quality of the synthesized thin films, and therefore neglecting the morphological differences, the chemical composition-induced changes in elastic modulus of $VN_x$ ($0.77 \leq x \leq 0.97$) are evaluated using nanoindentation experiments and the results are depicted in Figure 8 (b). Additionally, the calculated elastic moduli of $VN_x$ ($0.72 \leq x \leq 1.00$) are included, together with the measured and calculated elastic modulus of the stoichiometric $VN_{1.00}$ from literature. For the near-stoichiometric $VN_{0.97}$ thin film, the measured elastic modulus is 400 ± 25 GPa, which is in good agreement with other reported values for bulk [85], thin film [15], and calculated $VN_{1.00}$ [86]. As $x$ is reduced from 0.97 to 0.77, the elastic modulus increases continuously to 444 ± 24 GPa. This anomalous 11% increase in the stiffness of $VN_x$ due to the N vacancy presence is consistent with the predicted elasticities, as evident in Figure 8 (b). As shown in Figure 1 and Figure 4, this anomalous elasticity increase can be rationalized by the N vacancy-induced elastic anisotropy and bond strengthening, as a characteristic of the group IV binary cubic $VN_x$ compound. Based on the above presented theoretical predictions on the effect of VEC and N-vacancies on the elastic properties of group IV ($TiN_x$) and V ($VN_x$) cubic transition metal nitrides the different trends in the mechanical properties reported in literature can readily be rationalized: On one hand, experimental reports on N vacancy-induced reduction in elasticity of $TiN_x$ by Jhi *et al.* [33] and Shin *et al.* [31] are consistent with the here presented predictions. On the other hand, also the reported elasticity of the group V $NbN_x$ thin films exhibited an anomalous behavior with respect to N vacancy concentration [16].

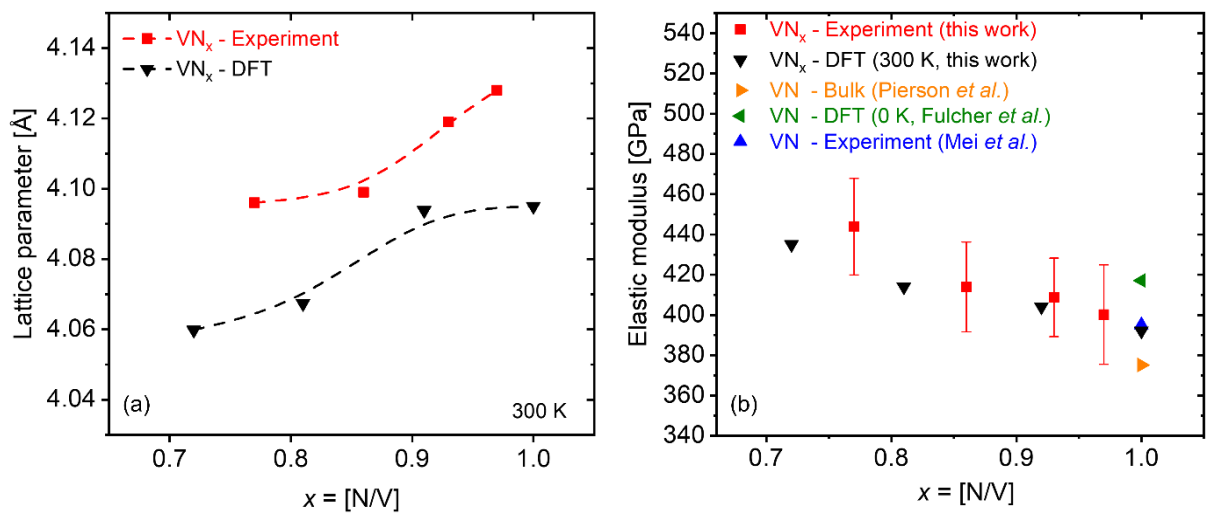

Figure 8. Calculated and measured (a) relaxed lattice parameters $a_0$ and (b) elastic moduli of $VN_x$ as a function of N/V ratio x in this work together with the elastic modulus values for bulk $VN_{1.00}$ (Pierson et al. [9]), DFT $VN_{1.00}$ at 0 K (Fulcher et al. [86]), and thin film $VN_{1.00}$ (Mei et al. [15]).

**Conclusions**

By employing *ab initio* calculations, we have systematically investigated the effect of N vacancy concentration on the elastic properties of cubic early transition metal nitrides, TiN$_x$, ZrN$_x$, VN$_x$, NbN$_x$, and MoN$_x$ ($0.72 \leq x \leq 1.00$). A different elastic response due to the presence of N vacancies is exhibited between the group IV, V, and VI nitrides and hence between different transition metal valence electron concentrations. In group IV TiN$_x$ and ZrN$_x$, N vacancies reduce the elastic modulus, while an opposite trend is observed for group V VN$_x$ and NbN$_x$, where an anomalous enhancement in elastic modulus is evident due to presence of N vacancies. To understand the origin of N vacancy-induced elastic response for different transition metal valence electron concentrations in these systems, the directional elastic modulus and bond characteristics within TiN$_x$ and VN$_x$ are compared. In TiN$_x$, N vacancy incorporation results in a significant reduction of elasticity along the [001], [011], and [111] directions. However, the vacancy-induced softening in elastic modulus for VN$_x$ along [001] is overcompensated by strengthening along [011] and [111] resulting in an increased elastic modulus. In addition, individual bond strengths and their averages are obtained from crystal orbital Hamilton population (COHP) analyses. By considering the integrated COHP values, directly correlated with the bond strength, N vacancies seem to weaken the Ti–N bond overall, with ICOHP values from –2.97 eV for TiN to –2.90 eV for TiN$_{0.72}$. In contrast, V–N bonds are strengthened, with ICOHP values from –2.62 for VN to –2.70 eV for VN$_{0.72}$ due to structural relaxation in the presence of N vacancies. To validate the anomalous elasticity enhancement with respect to N vacancy concentration experimentally, close-to-single-crystal VN$_x$ thin films ($0.77 \leq x \leq 0.97$) were grown on single crystal MgO(001) substrates. Reduction of N content in VN$_x$/MgO(001) from $x = 0.97 \pm 0.06$ to $0.77 \pm 0.04$ leads to a decrease in the relaxed lattice parameter $a_0$ from 4.128 Å to 4.096 Å, respectively, as determined by X-ray diffraction. The understoichiometric thin films show high in-plane coherence length $\xi_\parallel$ of up to 92 nm. Nanoindentation results reveal that the reduction in lattice parameter of $VN_x$ with increasing N vacancy concentration goes in hand with an anomalous 11% increase in elastic modulus and is in very good agreement with DFT predictions. Based on the theoretical and experimental data presented here, it is evident that the elastic behavior of early transition metal nitrides is defined by the N vacancy concentration and the transition metal valence electron concentration. COHP analyses and elastic anisotropy calculations reveal that this behavior can be rationalized by considering the direction-dependent bond-strengthening (bond-softening) upon vacancy introduction into $VN_x$ ($TiN_x$). This study offers a strategy to design nitride thin films with defined elastic properties by controlling the valence electron- and vacancy-concentration and is expected to be relevant for other isostructural binary and ternary nitrides as well.

**Acknowledgement**

This work was supported by the Deutsche Forschungsgemeinschaft (DFG) within the Collaborative Research Center SFB-TR 87/3 "Pulsed high power plasmas for the synthesis of nanostructured functional layers". Simulations were performed with computing resources granted by the Jülich-Aachen Research Alliance (JARA) HPC section at the IT Center of RWTH Aachen University under the projects JARA0151 and JARA0221 and their support is gratefully acknowledged. The authors also acknowledge financial support from the Swedish research council, VR-RFI (contract#2017-00646_9) and the Swedish Foundation for Strategic Research (SSF, contract RIF14-0053) supporting the operation of the tandem accelerator at Uppsala University. The authors also acknowledge the Swedish Research Council (VR) under project number 2021-03826.

**References**

[1] S. Koseki, K. Inoue, S. Morito, T. Ohba, H. Usuki, Comparison of TiN-coated tools using CVD and PVD processes during continuous cutting of Ni-based superalloys, Surface and Coatings Technology 283 (2015) 353-363.
[2] P. Hones, M. Diserens, R. Sanjinés, F. Lévy, Electronic structure and mechanical properties of hard coatings from the chromium–tungsten nitride system, Journal of Vacuum Science & Technology B: Microelectronics and Nanometer Structures Processing, Measurement, and Phenomena 18(6) (2000) 2851-2856.
[3] X.-P. Qu, M. Zhou, T. Chen, Q. Xie, G.-P. Ru, B.-Z. Li, Study of ultrathin vanadium nitride as diffusion barrier for copper interconnect, Microelectronic Engineering 83(2) (2006) 236-240.
[4] M. Mühlbacher, A.S. Bochkarev, F. Mendez-Martin, B. Sartory, L. Chitu, M.N. Popov, P. Puschnig, J. Spitaler, H. Ding, N. Schalk, J. Lu, L. Hultman, C. Mitterer, Cu diffusion in single-crystal and polycrystalline TiN barrier layers: A high-resolution experimental study supported by first-principles calculations, Journal of Applied Physics 118(8) (2015) 085307.
[5] L. Hultman, Thermal stability of nitride thin films, Vacuum 57(1) (2000) 1-30.
[6] J.C. Caicedo, G. Zambrano, W. Aperador, L. Escobar-Alarcon, E. Camps, Mechanical and electrochemical characterization of vanadium nitride (VN) thin films, Applied Surface Science 258(1) (2011) 312-320.
[7] J. Musil, Hard and superhard nanocomposite coatings, Surface and Coatings Technology 125(1) (2000) 322-330.
[8] E.E. Vera, M. Vite, R. Lewis, E.A. Gallardo, J.R. Laguna-Camacho, A study of the wear performance of TiN, CrN and WC/C coatings on different steel substrates, Wear 271(9) (2011) 2116-2124.
[9] W.D. Münz, Titanium aluminum nitride films: A new alternative to TiN coatings, Journal of Vacuum Science & Technology A 4(6) (1986) 2717-2725.
[10] P.H. Mayrhofer, A. Hörling, L. Karlsson, J. Sjölén, T. Larsson, C. Mitterer, L. Hultman, Self-organized nanostructures in the Ti–Al–N system, Applied Physics Letters 83(10) (2003) 2049-2051.
[11] D.M. Holzapfel, D. Music, M. Hans, S. Wolff-Goodrich, D. Holec, D. Bogdanovski, M. Arndt, A.O. Eriksson, K. Yalamanchili, D. Primetzhofer, C.H. Liebscher, J.M. Schneider, Enhanced thermal stability of (Ti,Al)N coatings by oxygen incorporation, Acta Materialia 218 (2021) 117204.
[12] J.E. Sundgren, Structure and properties of TiN coatings, Thin Solid Films 128(1) (1985) 21-44.
[13] A.B. Mei, B.M. Howe, C. Zhang, M. Sardela, J.N. Eckstein, L. Hultman, A. Rockett, I. Petrov, J.E. Greene, Physical properties of epitaxial ZrN/MgO(001) layers grown by reactive magnetron sputtering, Journal of Vacuum Science & Technology A 31(6) (2013) 061516.
[14] H.-S. Seo, T.-Y. Lee, J.G. Wen, I. Petrov, J.E. Greene, D. Gall, Growth and physical properties of epitaxial HfN layers on MgO(001), Journal of Applied Physics 96(1) (2004) 878-884.
[15] A.B. Mei, R.B. Wilson, D. Li, D.G. Cahill, A. Rockett, J. Birch, L. Hultman, J.E. Greene, I. Petrov, Elastic constants, Poisson ratios, and the elastic anisotropy of VN(001), (011), and (111) epitaxial layers grown by reactive magnetron sputter deposition, Journal of Applied Physics 115(21) (2014) 214908.
[16] K. Zhang, K. Balasubramanian, B.D. Ozsdolay, C.P. Mulligan, S.V. Khare, W.T. Zheng, D. Gall, Growth and mechanical properties of epitaxial NbN(001) films on MgO(001), Surface and Coatings Technology 288 (2016) 105-114.
[17] C.-S. Shin, D. Gall, P. Desjardins, A. Vailionis, H. Kim, I. Petrov, J.E. Greene, M. Odén, Growth and physical properties of epitaxial metastable cubic TaN(001), Applied Physics Letters 75(24) (1999) 3808-3810.
[18] D. Gall, C.-S. Shin, T. Spila, M. Odén, M.J.H. Senna, J.E. Greene, I. Petrov, Growth of single-crystal CrN on MgO(001): Effects of low-energy ion-irradiation on surface morphological evolution and physical properties, Journal of Applied Physics 91(6) (2002) 3589-3597.
[19] B.D. Ozsdolay, K. Balasubramanian, D. Gall, Cation and anion vacancies in cubic molybdenum nitride, Journal of Alloys and Compounds 705 (2017) 631-637.
[20] B.D. Ozsdolay, C.P. Mulligan, K. Balasubramanian, L. Huang, S.V. Khare, D. Gall, Cubic β-WN$_x$ layers: Growth and properties vs N-to-W ratio, Surface and Coatings Technology 304 (2016) 98-107.

[21] D.G. Sangiovanni, B. Alling, P. Steneteg, L. Hultman, I.A. Abrikosov, Nitrogen vacancy, self-interstitial diffusion, and Frenkel-pair formation/dissociation in *B1* TiN studied by *ab initio* and classical molecular dynamics with optimized potentials, Physical Review B 91(5) (2015) 054301.
[22] K. Balasubramanian, S.V. Khare, D. Gall, Energetics of point defects in rocksalt structure transition metal nitrides: Thermodynamic reasons for deviations from stoichiometry, Acta Materialia 159 (2018) 77-88.
[23] S. Karimi Aghda, D. Music, Y. Unutulmazsoy, H.H. Sua, S. Mráz, M. Hans, D. Primetzhofer, A. Anders, J.M. Schneider, Unravelling the ion-energy-dependent structure evolution and its implications for the elastic properties of (V,Al)N thin films, Acta Materialia 214 (2021) 117003.
[24] D.M. Holzapfel, D. Music, S. Mráz, S.K. Aghda, M. Etter, P. Ondračka, M. Hans, D. Bogdanovski, S. Evertz, L. Patterer, P. Schmidt, A. Schökel, A.O. Eriksson, M. Arndt, D. Primetzhofer, J.M. Schneider, Influence of ion irradiation-induced defects on phase formation and thermal stability of $Ti_{0.27}Al_{0.21}N_{0.52}$ coatings, Acta Materialia (2022) 118160.
[25] L. Tsetseris, N. Kalfagiannis, S. Logothetidis, S.T. Pantelides, Structure and interaction of point defects in transition-metal nitrides, Physical Review B 76(22) (2007) 224107.
[26] D. Music, L. Banko, H. Ruess, M. Engels, A. Hecimovic, D. Grochla, D. Rogalla, T. Brögelmann, A. Ludwig, A.v. Keudell, K. Bobzin, J.M. Schneider, Correlative plasma-surface model for metastable Cr-Al-N: Frenkel pair formation and influence of the stress state on the elastic properties, Journal of Applied Physics 121(21) (2017) 215108.
[27] N. Koutná, D. Holec, O. Svoboda, F.F. Klimashin, P.H. Mayrhofer, Point defects stabilise cubic Mo-N and Ta-N, Journal of Physics D: Applied Physics 49(37) (2016) 375303.
[28] K. Balasubramanian, L. Huang, D. Gall, Phase stability and mechanical properties of $Mo_{1-x}N_x$ with $0 \leqslant x \leqslant 1$, Journal of Applied Physics 122(19) (2017) 195101.
[29] K. Balasubramanian, S. Khare, D. Gall, Vacancy-induced mechanical stabilization of cubic tungsten nitride, Physical Review B 94(17) (2016) 174111.
[30] T. Lee, K. Ohmori, C.S. Shin, D.G. Cahill, I. Petrov, J.E. Greene, Elastic constants of single-crystal $TiN_x(001)$ ($0.67 \leqslant x \leqslant 1.0$) determined as a function of *x* by picosecond ultrasonic measurements, Physical Review B 71(14) (2005) 144106.
[31] C.-S. Shin, D. Gall, N. Hellgren, J. Patscheider, I. Petrov, J.E. Greene, Vacancy hardening in single-crystal TiNx(001) layers, Journal of Applied Physics 93(10) (2003) 6025-6028.
[32] H. Kindlund, D.G. Sangiovanni, J. Lu, J. Jensen, V. Chirita, J. Birch, I. Petrov, J.E. Greene, L. Hultman, Vacancy-induced toughening in hard single-crystal $V_{0.5}Mo_{0.5}N_x/MgO(001)$ thin films, Acta Materialia 77 (2014) 394-400.
[33] S.H. Jhi, S.G. Louie, M.L. Cohen, J. Ihm, Vacancy hardening and softening in transition metal carbides and nitrides, Phys Rev Lett 86(15) (2001) 3348-51.
[34] H.-S. Seo, T.-Y. Lee, I. Petrov, J.E. Greene, D. Gall, Epitaxial and polycrystalline $HfN_x$ ($0.8 \leqslant x \leqslant 1.5$) layers on MgO(001): Film growth and physical properties, Journal of Applied Physics 97(8) (2005) 083521.
[35] M. Benkahoul, E. Martinez, A. Karimi, R. Sanjinés, F. Lévy, Structural and mechanical properties of sputtered cubic and hexagonal $NbN_x$ thin films, Surface and Coatings Technology 180-181 (2004) 178-183.
[36] C.-S. Shin, D. Gall, Y.-W. Kim, P. Desjardins, I. Petrov, J.E. Greene, M. Odén, L. Hultman, Epitaxial NaCl structure δ-$TaN_x$(001): Electronic transport properties, elastic modulus, and hardness versus N/Ta ratio, Journal of Applied Physics 90(6) (2001) 2879-2885.
[37] D. Gall, I. Petrov, N. Hellgren, L. Hultman, J.E. Sundgren, J.E. Greene, Growth of poly- and single-crystal ScN on MgO(001): Role of low-energy $N_2^+$ irradiation in determining texture, microstructure evolution, and mechanical properties, Journal of Applied Physics 84(11) (1998) 6034-6041.
[38] C.-S. Shin, S. Rudenja, D. Gall, N. Hellgren, T.-Y. Lee, I. Petrov, J.E. Greene, Growth, surface morphology, and electrical resistivity of fully strained substoichiometric epitaxial $TiN_x$ ($0.67 \leqslant x < 1.0$) layers on MgO(001), Journal of Applied Physics 95(1) (2004) 356-362.
[39] A.B. Mei, M. Tuteja, D.G. Sangiovanni, R.T. Haasch, A. Rockett, L. Hultman, I. Petrov, J.E. Greene, Growth, nanostructure, and optical properties of epitaxial $VN_x/MgO(001)$ ($0.80 \leqslant x \leqslant 1.00$) layers deposited by reactive magnetron sputtering, Journal of Materials Chemistry C 4(34) (2016) 7924-7938.
[40] P. Hohenberg, W. Kohn, Inhomogeneous Electron Gas, Physical Review 136(3B) (1964) B864-B871.

[41] W. Kohn, L.J. Sham, Self-Consistent Equations Including Exchange and Correlation Effects, Physical Review 140(4A) (1965) A1133-A1138.
[42] G. Kresse, J. Hafner, Ab initio molecular-dynamics simulation of the liquid-metal--amorphous-semiconductor transition in germanium, Physical Review B 49(20) (1994) 14251-14269.
[43] G. Kresse, J. Furthmüller, Efficient iterative schemes for ab initio total-energy calculations using a plane-wave basis set, Physical Review B 54(16) (1996) 11169-11186.
[44] G. Kresse, J. Furthmüller, Efficiency of ab-initio total energy calculations for metals and semiconductors using a plane-wave basis set, Computational Materials Science 6(1) (1996) 15-50.
[45] P.E. Blöchl, Projector augmented-wave method, Physical Review B 50(24) (1994) 17953-17979.
[46] G. Kresse, D. Joubert, From ultrasoft pseudopotentials to the projector augmented-wave method, Physical Review B 59(3) (1999) 1758-1775.
[47] J.P. Perdew, K. Burke, M. Ernzerhof, Generalized Gradient Approximation Made Simple, Physical Review Letters 77(18) (1996) 3865-3868.
[48] H.J. Monkhorst, J.D. Pack, Special points for Brillouin-zone integrations, Physical Review B 13(12) (1976) 5188-5192.
[49] M. Methfessel, A.T. Paxton, High-precision sampling for Brillouin-zone integration in metals, Physical Review B 40(6) (1989) 3616-3621.
[50] R. Yu, J. Zhu, H.Q. Ye, Calculations of single-crystal elastic constants made simple, Computer Physics Communications 181(3) (2010) 671-675.
[51] M. Moakher, A.N. Norris, The Closest Elastic Tensor of Arbitrary Symmetry to an Elasticity Tensor of Lower Symmetry, Journal of Elasticity 85(3) (2006) 215-263.
[52] R. Hill, The Elastic Behaviour of a Crystalline Aggregate, Proceedings of the Physical Society. Section A 65(5) (1952) 349-354.
[53] P. Söderlind, L. Nordström, Y. Lou, B. Johansson, Relativistic effects on the thermal expansion of the actinide elements, Physical Review B 42(7) (1990) 4544-4552.
[54] H. Rueß, D. Music, A. Bahr, J.M. Schneider, Effect of chemical composition, defect structure, and stress state on the elastic properties of $(V_{1-x}Al_x)_{1-y}N_y$, Journal of Physics: Condensed Matter 32(2) (2019) 025901.
[55] J.F. Nye, Physical properties of crystals: their representation by tensors and matrices, Oxford university press1985.
[56] P.E. Blöchl, O. Jepsen, O.K. Andersen, Improved tetrahedron method for Brillouin-zone integrations, Physical Review B 49(23) (1994) 16223-16233.
[57] V.L. Deringer, A.L. Tchougréeff, R. Dronskowski, Crystal Orbital Hamilton Population (COHP) Analysis As Projected from Plane-Wave Basis Sets, The Journal of Physical Chemistry A 115(21) (2011) 5461-5466.
[58] S. Maintz, V.L. Deringer, A.L. Tchougréeff, R. Dronskowski, Analytic projection from plane-wave and PAW wavefunctions and application to chemical-bonding analysis in solids, Journal of Computational Chemistry 34(29) (2013) 2557-2567.
[59] S. Maintz, V.L. Deringer, A.L. Tchougréeff, R. Dronskowski, LOBSTER: A tool to extract chemical bonding from plane-wave based DFT, Journal of Computational Chemistry 37(11) (2016) 1030-1035.
[60] R. Nelson, C. Ertural, J. George, V.L. Deringer, G. Hautier, R. Dronskowski, LOBSTER: Local orbital projections, atomic charges, and chemical-bonding analysis from projector-augmented-wave-based density-functional theory, Journal of Computational Chemistry 41(21) (2020) 1931-1940.
[61] R. Dronskowski, P.E. Bloechl, Crystal orbital Hamilton populations (COHP): energy-resolved visualization of chemical bonding in solids based on density-functional calculations, The Journal of Physical Chemistry 97(33) (1993) 8617-8624.
[62] R. Dronskowski, Computational Chemistry of Solid State Materials a quide for materials scientists, chemists, physicists and others, Wiley-Blackwell, Weinheim2005.
[63] G.A. Landrum, R. Dronskowski, The Orbital Origins of Magnetism: From Atoms to Molecules to Ferromagnetic Alloys, Angewandte Chemie International Edition 39(9) (2000) 1560-1585.
[64] S. Amano, D. Bogdanovski, H. Yamane, M. Terauchi, R. Dronskowski, ε-TiO, a Novel Stable Polymorph of Titanium Monoxide, Angewandte Chemie International Edition 55(5) (2016) 1652-1657.
[65] D. Bogdanovski, P.J. Pöllmann, J.M. Schneider, An ab initio investigation of the temperature-dependent energetic barriers towards CrAlB and (Mo,Cr)AlB formation in a metastable synthesis scenario, Nanoscale 14(35) (2022) 12866-12874.

[66] H. Windischmann, Intrinsic stress in sputter-deposited thin films, Critical Reviews in Solid State and Materials Sciences 17(6) (1992) 547-596.
[67] S. Karimi Aghda, D.M. Holzapfel, D. Music, Y. Unutulmazsoy, S. Mráz, D. Bogdanovski, G. Fidanboy, M. Hans, D. Primetzhofer, A.S.J. Méndez, A. Anders, J.M. Schneider, Ion kinetic energy- and ion flux-dependent mechanical properties and thermal stability of (Ti,Al)N thin films, Acta Materialia 250 (2023) 118864.
[68] I. Petrov, V. Orlinov, I. Ivanov, J. Kourtev, Electrostatic Probe Measurements in the Glow Discharge Plasma of a D. C. Magnetron Sputtering System, Contributions to Plasma Physics 28(2) (1988) 157-167.
[69] J.J. Arnaud Le Febvrier, Per Eklund, Wet-cleaning of MgO(001): Modification of surface chemistry and effects on thin film growth investigated by x-ray photoelectron spectroscopy and time-of-flight secondary ion mass spectroscopy, Journal of Vacuum Science & Technology A 35 (2017).
[70] P. Ström, D. Primetzhofer, Ion beam tools for nondestructive in-situ and in-operando composition analysis and modification of materials at the Tandem Laboratory in Uppsala, Journal of Instrumentation 17(04) (2022) P04011.
[71] M. Janson, CONTES Conversion of Time-energy Spectra a Program for ERDA Data Analysis (Internal Report, Uppsala University), Uppsala, 2004.
[72] M.A. Sortica, V. Paneta, B. Bruckner, S. Lohmann, M. Hans, T. Nyberg, P. Bauer, D. Primetzhofer, Electronic energy-loss mechanisms for H, He, and Ne in TiN, Physical Review A 96(3) (2017) 032703.
[73] S.M. Durbin, J.E. Cunningham, C.P. Flynn, Growth of single-crystal metal superlattices in chosen orientations, Journal of Physics F: Metal Physics 12(6) (1982) L75.
[74] R.C. Powell, N.E. Lee, Y.W. Kim, J.E. Greene, Heteroepitaxial wurtzite and zinc‐blende structure GaN grown by reactive‐ion molecular‐beam epitaxy: Growth kinetics, microstructure, and properties, Journal of Applied Physics 73(1) (1993) 189-204.
[75] S. Zak, C.O.W. Trost, P. Kreiml, M.J. Cordill, Accurate measurement of thin film mechanical properties using nanoindentation, Journal of Materials Research 37(7) (2022) 1373-1389.
[76] W.C. Oliver, G.M. Pharr, An improved technique for determining hardness and elastic modulus using load and displacement sensing indentation experiments, Journal of Materials Research 7(6) (2011) 1564-1583.
[77] D. Holec, M. Friák, J. Neugebauer, P.H. Mayrhofer, Trends in the elastic response of binary early transition metal nitrides, Physical Review B 85(6) (2012) 064101.
[78] C. Ravi, H.K. Sahu, M.C. Valsakumar, A. van de Walle, Cluster expansion Monte Carlo study of phase stability of vanadium nitrides, Physical Review B 81(10) (2010) 104111.
[79] D.G. Sangiovanni, A.B. Mei, L. Hultman, V. Chirita, I. Petrov, J.E. Greene, Ab Initio Molecular Dynamics Simulations of Nitrogen/VN(001) Surface Reactions: Vacancy-Catalyzed $N_2$ Dissociative Chemisorption, N Adatom Migration, and $N_2$ Desorption, The Journal of Physical Chemistry C 120(23) (2016) 12503-12516.
[80] M.T. A. B. Mei, D. G. Sangiovanni, R. T. Haasch, A. Rockett, L. Hultman, I. Petrov and J. E. Greene, Growth, nanostructure, and optical properties of epitaxial $VN_x$/MgO(001) (0.80 r x r 1.00) layers deposited by reactive magnetron sputtering, Journal of Materials Chemistry C (2016).
[81] R. Sanjinés, C. Wiemer, P. Hones, F. Lévy, Chemical bonding and electronic structure in binary $VN_y$ and ternary $T_{1-x}V_xN_y$ nitrides, Journal of Applied Physics 83(3) (1998) 1396-1402.
[82] L. Porte, L. Roux, J. Hanus, Vacancy effects in the x-ray photoelectron spectra of $TiN_x$, Physical Review B 28(6) (1983) 3214-3224.
[83] R. Sanjinés, C. Wiemer, J. Almeida, F. Lévy, Valence band photoemission study of the Ti-Mo-N system, Thin Solid Films 290-291 (1996) 334-338.
[84] I. Petrov, P.B. Barna, L. Hultman, J.E. Greene, Microstructural evolution during film growth, Journal of Vacuum Science & Technology A 21(5) (2003) S117-S128.
[85] H.O. Pierson, 11 - Interstitial Nitrides: Properties and General Characteristics, in: H.O. Pierson (Ed.), Handbook of Refractory Carbides and Nitrides, William Andrew Publishing, Westwood, NJ, 1996, pp. 181-208.
[86] B.D. Fulcher, X.Y. Cui, B. Delley, C. Stampfl, Hardness analysis of cubic metal mononitrides from first principles, Physical Review B 85(18) (2012) 184106.

**Supplementary materials**

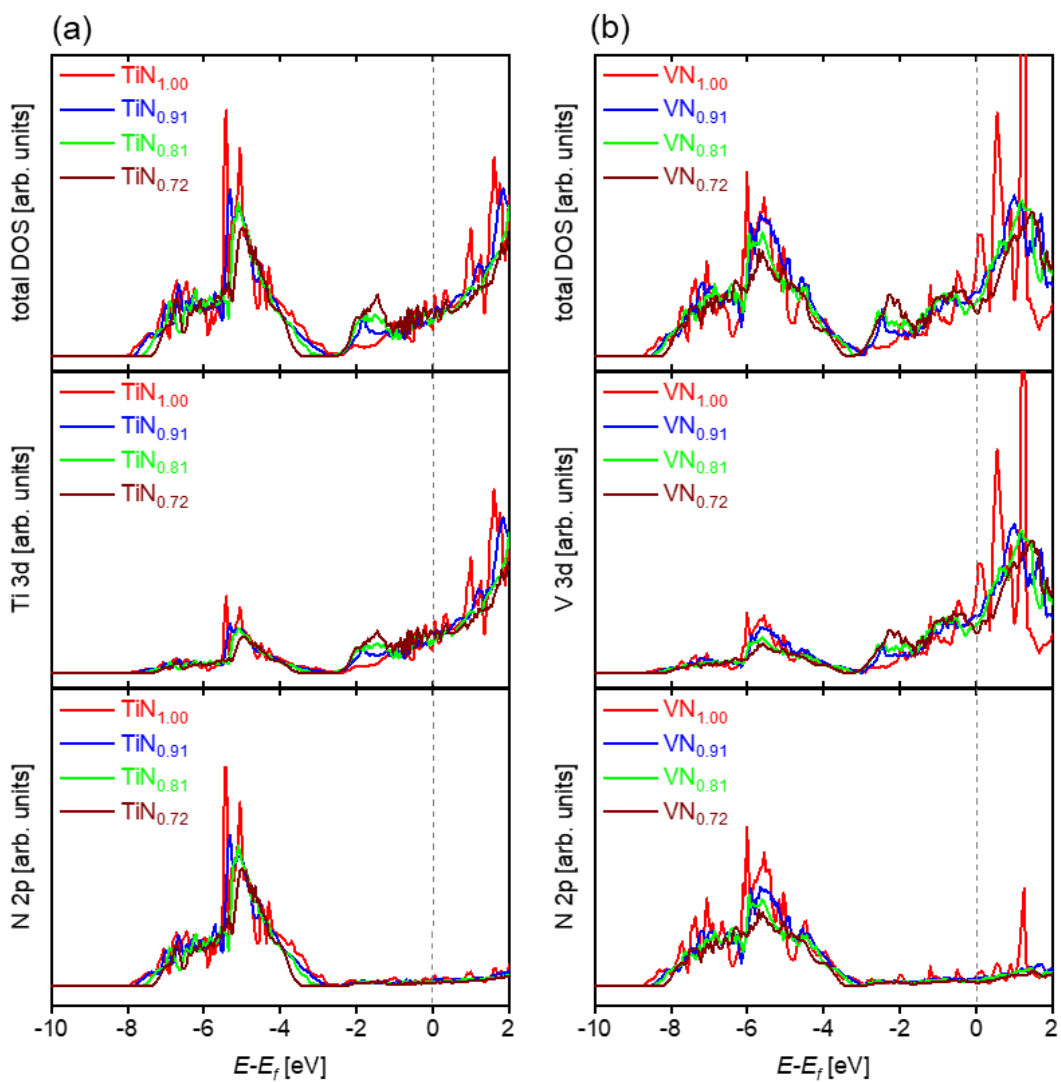

Figure S 1. Total and partial density of states (DOS) analysis for (a) $TiN_x$ and (b) $VN_x$ as a function of x. $E_f$ designates the Fermi energy.